\documentstyle[prl,aps,multicol,epsf]{revtex}
\begin{document}
\draft
\title{Microwave Rectification at the Boundary between
Two-Dimensional Electron Systems.}
\author{I. Hoxha, S. A. Vitkalov, N. A. Zimbovskaya, and M.P. Sarachik}
\address{Physics Department, City College of the City
University of New York, New York, New York 10031}
\author{T.~M.~Klapwijk}
\address{Delft University of Technology, Department of Applied Physics,
2628 CJ Delft, The Netherlands}
\date{\today}
\maketitle

\begin{abstract}

Rectification of microwave radiation (20-40 GHz) by a line boundary
between two two-dimensional metals on a silicon surface was observed
and investigated at different temperatures, in-plane magnetic fields and
microwave powers. The rectified voltage $V_{dc}$ is generated whenever
the electron densities $n_{1,2}$ of the two metals are different,
changing polarity at $n_1 \approx n_2$.  Very strong nonlinear response is
found when one of the two 2D metals is close to the electron density
corresponding to the reported magnetic instability in this system.

\end{abstract}


\begin{multicols}{2}

\section{Introduction}

Dilute low dimensional systems have been the focus of a great deal of
recent attention due to their interesting physical properties at low
temperature\cite{Abrahams}.  A number of fascinating phenomena have been
reported for low-density electron systems in high mobility silicon
inversion layers as a function of magnetic field.  A dramatic increase
of the resistivity in response to in-plane magnetic field has been shown
to be associated at high electron density with complete spin
polarization of the carriers \cite{okamoto,vitkalov,vitkalovangular}.
A number of experiments have recently shown that the spin susceptibility
increases substantially as the density is decreased
\cite{okamoto,Shashkin,pudalov_SdH}, indicating a possible divergence
and ferromagnetic instability at finite electron density.  Through a detailed
study of the magnetoconductivity as a function of temperature and electron
density, we have identified an energy scale $\Delta$ associated with the
response of the electrons to a magnetic field applied parallel to the plane of
the electrons
\cite{vitkalov_ferro}.  The density-dependent energy $\Delta$ was found
to go to zero at a finite density $n_0$, signaling critical behavior and
the occurence of a zero-temperature quantum phase transition of magnetic
origin.

Microwave radiation may be an interesting tool for probing the behavior
of these systems.  At low frequencies $\hbar \omega \ll \Delta$ the 
system is expected to behave as a correlated spin liquid, while at
high frequencies $\hbar \omega \gg \Delta$ the behavior will be
similar to that of an noninteracting gas.  In response to microwave
radiation of frequency $\omega$, one may find particularly interesting
behavior at the boundary between two metallic regions which have
characteristic energies $\Delta_1$ and $ \Delta_2$ if $\Delta_1 < \hbar
\omega < \Delta_2$, corresponding to transport between a correlated spin
liquid and a noninteracting gas.

In this  paper we report unusual high frequency behavior of dilute 2D
electron systems in multigated high mobility Si-MOSFET's.  We
investigated the nonlinear microwave response of the boundary between two
regions of the silicon inversion layers with different electron
densities.  A rectified signal is observed whenever the electron
densities $n_{1,2}$ of the two metals are different.  The polarity of
the rectified signal can be changed easily by varying the electron
density on the two sides of the boundary.  We find that the nonlinear
response of the boundary between the two metals is unusually strong when
one of the metals is close to the quantum phase transition reported
earlier \cite{vitkalov_ferro} while the other is kept at high electron
density - a condition which corresponds to $\Delta_1 < \hbar \omega <
\Delta_2$.

\section{Experimental Setup}
	
   We used multigated Si-MOSFET's with six different contacts,
shown schematically in Fig. 1.  The 2D electron densities $n_i$ corresponding 
to different contacts were separately controlled by independent gates (contact
gates).  Each 2D metal system was adjacent to a common 2D conducting
region of density $n_2$, also controlled independently by a main gate. 
The region between the main gate and each contact gate was in the form
of a thin-line split. The narrow split between gate metallizations
was obtained by reactive ion etching\cite{heemskerk}.  The typical width
of a split ($50-70$ nm) was less than the thickness of the Si oxide
insulating layer ($152$ nm), providing a smooth, probably monotonic,
profile of electron density between two 2D layers with different gate
voltages, as shown in Fig. 1 (c).  The orientation of two splits, shown
as vertical lines on the surface of the structure in Fig. 1(a), was
perpendicular to the orientation of the remaining four horizontal
splits.  The two contacts corresponding to the vertical splits  were
permanently connected to ground. 
\vbox{
\vspace{0 in}
\hbox{
\hspace{+0.1in} 
\epsfxsize 3.4 in \epsfbox{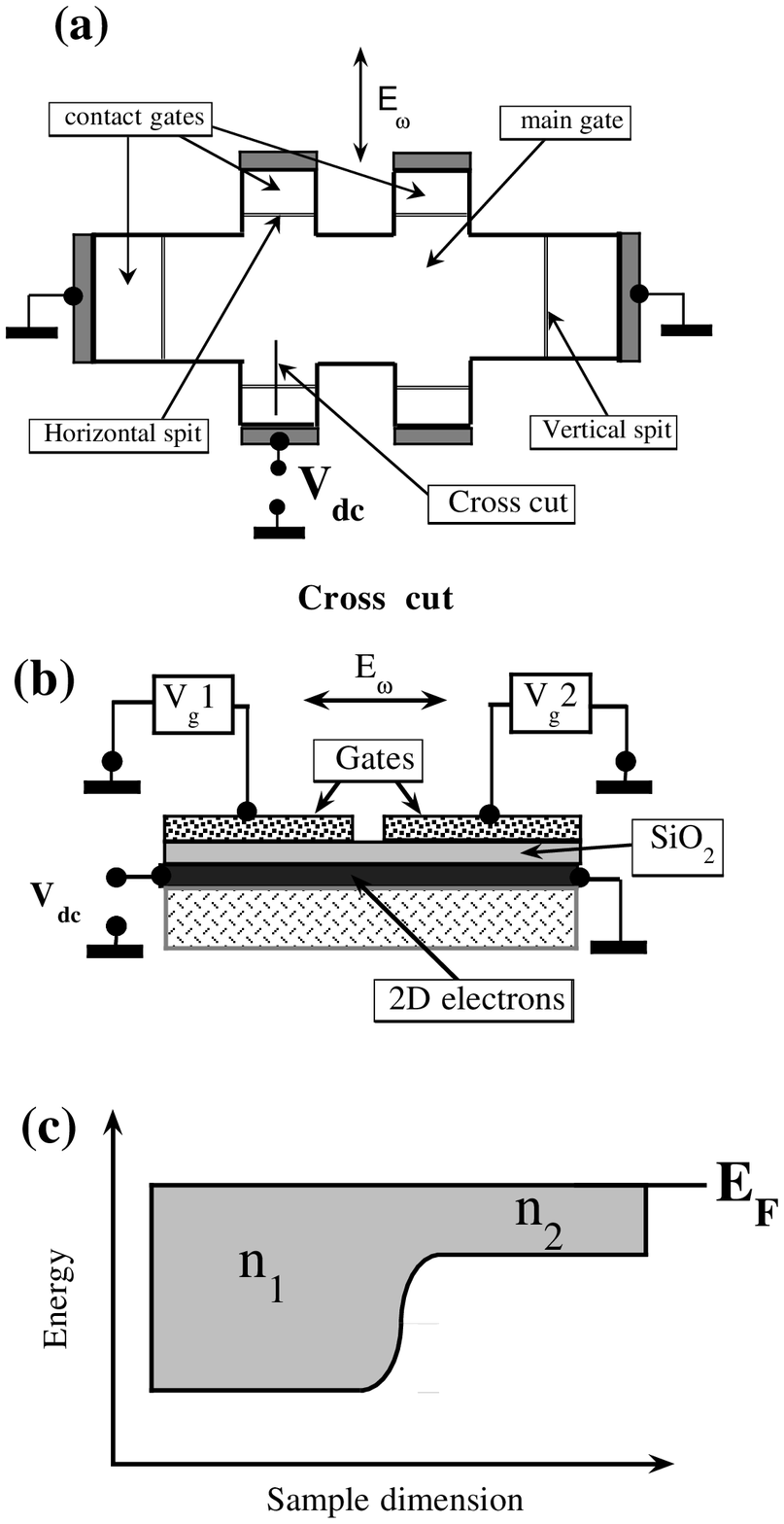} 
}
}
\vskip 0.5cm
\refstepcounter{figure}
\parbox[b]{3.3in}{\baselineskip=12pt FIG.~\thefigure.
(a) Top view of the sample.
(b) Cross-sectional view of the sample between one contact area and the
main sample area.  The regions under the two gates form
two different 2D metals (2D-Metal \#1 and 2D-Metal \#2). 
(c) the Fermi level and bottom of the conduction band are shown
as a function of position along the sample. The shaded area below the
Fermi level corresponds to occupied electron states of the two 2D metals
with different electron densities $n_1$ and $n_2$.
\vspace{0.10in}
}
\label{1} 
By varying the frequency of the
microwave radiation, which changes the configuration of the
electromagnetic field near the sample, it was possible to vary the
relative contributions of vertical and horizontal splits to the measured
nonlinear signal.  Measurements were taken at a frequency near $20$ GHz,
where the nonlinear contribution from the vertical splits
was negligibly small (less than $3-5$\%). Thus, the individual nonlinear
properties of each horizontal split were studied in the experiment. Fig.1
(b) shows a simplified diagram of the experimental setup near the
horizontal split; the two two-dimensional metals with electron density
controlled independently by gates \#$1$ and \#$2$ are connected through
the conducting region under the split.  The rectified signal, a DC
voltage $V_{dc}$, is observed in response to a microwave electric field
$E_\omega$ applied to the structure as shown in Fig.1. 
The voltage $V_{dc}$ was measured with respect to ground using a high
input impedance (10 G$\Omega$) voltmeter. 
\vbox{
\vspace{0.1 in}
\hbox{
\hspace{-0.0in} 
\epsfxsize 3.6 in \epsfbox{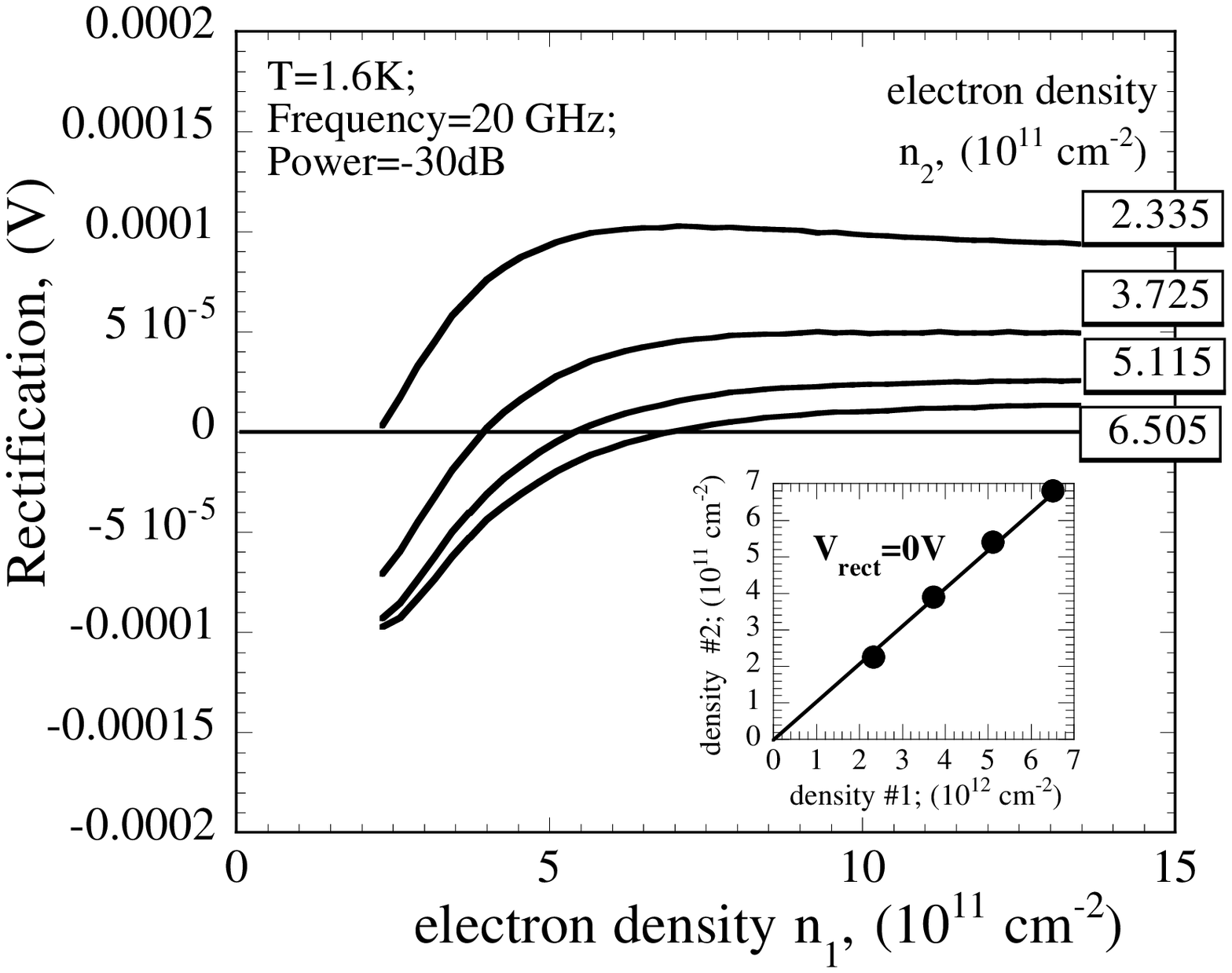} 
}
}
\vskip -3.5cm
\refstepcounter{figure}
\parbox[b]{3.3in}{\baselineskip=12pt FIG.~\thefigure.
Rectified signal versus electron density of 2D-metal \#1, with the
electron density of 2D-metal \#2 kept constant at the labeled values.
$T=1.6$ K, the frequency of radiation is $20$ GHz and the power
attenuation is $-30$ dB.  The insert shows the electron density of
2D-metal \#2 versus electron density of 2D-metal \#1 at which the
rectified signal changes polarity (goes through zero).
\vspace{0.10in}
}
\label{2}
DC currents flowing through the
system were thus exceedingly small, and the odd-order nonlinear
signals ($3, 5$...) of the different contacts provided negligible
contributions to the measured rectified signal.  Microwave radiation 
of frequency $20-40$ GHz was provided by a loop antenna
maintained at the end of a microwave coaxial line.  The sample was placed
near the radiation loop at a distance of about $1$ cm. The axis of the
microwave loop was approximately parallel to the vertical splits, shifted
from them by approximately $0.5$ cm along the sample plane. The sample
sizes are $480 \times 50 \mu$m$^2$.  A microwave YIG generator with
electronically controlled frequency supplied a regulated microwave power
of about $1$ mW ($0$ dB) at the end of the coaxial line.  The sample was
mounted on a rotating platform at the end of a low temperature He-$3$
probe.  Measurements were taken in an Oxford He-$3$ system in the
temperature range $0.25-12$ K for different microwave power in 
magnetic fields up to $12$ T applied parallel to the plane.   The
multigated Si MOSFET samples used in these experiments had electron
mobilities of about $2.5$ m$^2$/Vs at $0.3$ K.  Three different contacts
were investigated; all demonstrated similar non-linear properties.  
\vbox{
\vspace{-0.3 in}
\hbox{
\hspace{-0.1 in} 
\epsfxsize 3.4 in \epsfbox{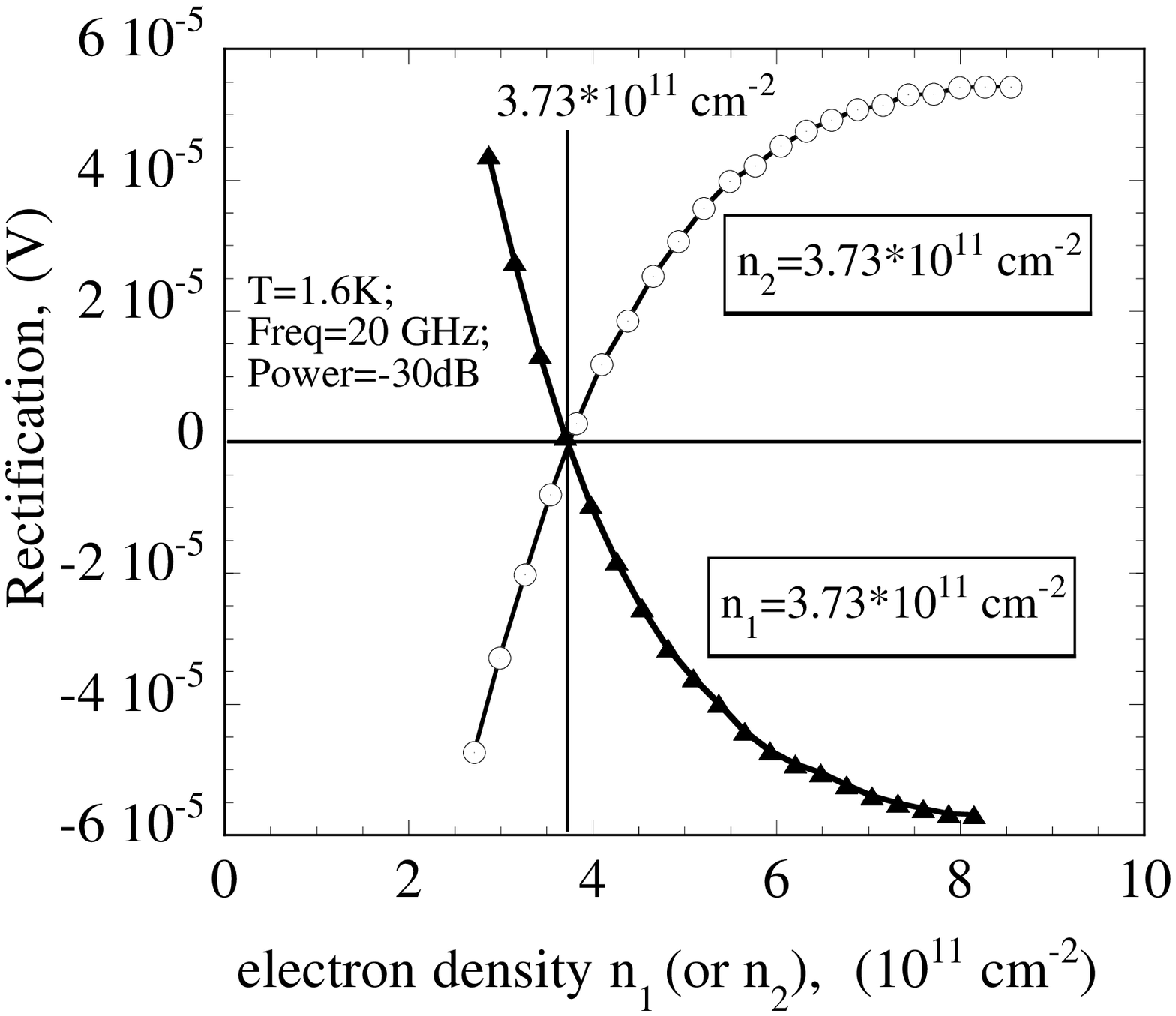} 
}
}
\vskip 0 cm
\refstepcounter{figure}
\parbox[b]{3.3in}{\baselineskip=12pt FIG.~\thefigure.
Open circles denote the rectified signal plotted as a function of electron
density of 2D-metal\#1 while the electron density of 2D-metal\#2 is kept
constant at $3.73x10^{11}$ cm$^-2$.  The triangles show the
rectified signal plotted versus electron density of 2D-metal\#2  when
electron density of 2D-metal\#1 is kept constant at $3.73x10^{11}$
cm$^2$.  For both curves the rectification is zero when the densities
of the two metals are equal.  Note that if $n_1$ and
$n_2$ are interchanged the rectification changes sign and is nearly
symmetric with respect to the horizontal axis. 
\vspace{0.10in}
}
\label{3} 

\section{Experimental Results}

\subsection{Density dependence} 

Fig. 2 shows the dependence of the rectified signal on the electron
density of two adjacent 2D metals.  The density of electrons in the
main 2D area is fixed by the main gate at some value, as labeled, while
the electron density in the contact area was varied.  The rectified
signal depends strongly on the electron density, changing polarity
when the electron densities of the two metals are nearly the same. The
insert shows the relation between the two densities $n_1$ and $n_2$ for
which the signal changes sign.  The correlation observed between the
electron densities on different sides of the boundary demonstrates that
the microwave rectification is generated by a region near to the boundary
between the two 2D metals; there is almost no rectification when the
two metals have the same electron density.  This implies that the
microwave rectification by the 2D metals \cite{Falko} is much smaller than
the rectification associated with the boundary between them.
The symmetry shown in Fig. 3 provides additional support for this
conclusion.  Fig. 3 demonstrates an inversion of the density dependence of
the rectification when the gates are interchanged.  Curve (a)
corresponds to measurements when the main gate voltage was fixed and
the contact gate voltage was varied while curve (b) shows the results
obtained for a fixed voltage on the contact gate when the main gate
voltage was varied.   The almost perfect symmetry observed
when the gates are interchanged demonstrates directly that the observed
rectification originates near the boundary between the two
two-dimensional metals.  

At fixed difference of electron densities of
the two metals the magnitude of the rectified signal decreases
considerably with increasing density (see fig. 2).  This decrease
indicates that the mismatch in physical properties responsible for the
nonlinearity at the boundary between the two metals becomes relatively 
small at high electron densities.

\subsection{Dependence of the rectification on microwave power}                                                           
The rectified signal exhibits different behavior as a function of
microwave power at low $(0.25$ K ) and high ($1.6$ K) temperatures.  The
nonlinear response is weaker at high temperature than at low
temperature.  

At high temperature (T=$1.6$ K) the rectified voltage is
proportional to the square of the microwave electric field $E_\omega^2$
at small microwave power ( $<$-30 dB), indicating that this is the weak
(perturbative) nonlinear regime (see Fig.4 (a)).  However, at
higher microwave excitation ($> -30$ dB) strongly nonlinear behavior is
observed: $V_{dc} \sim E_\omega$. 

At low temperature the rectified signal is roughly proportional to the
square root of the microwave power $V_{dc}\sim (P_\omega)^{1/2} \sim
E_\omega$ at all available powers if one of the two metals is kept at a
low electron density ($\sim 1 \times 10^{11}$ cm$^{-2}$), while the other
is held at high electron density (see Fig. 4 (b))\cite{high_dens}.  Such
strong nonlinearity of the conducting boundary between two metals is
unusual.  Estimates based on the Boltzmann equation (see below) show that
a weak, perturbative nonlinear response is expected over the entire range 
of microwave powers used in our experiments.  

The strong nonlinearity $V_{dc} \sim (P_\omega)^{1/2}$ of the boundary is 
similar to the strong nonlinear (high power) behavior of an ordinary diode
\cite{Levine}.  Using the analogy with a diode, we can estimate the
amplitude of the microwave voltage $V_\omega$ across the boundary.  The
current voltage characteristic of a diode is
\cite{Levine} 
$$I(V)=I_0 [e^{(eV/kT)}-1] \eqno{(1)},$$ 
where $I_0$ is the reverse current and $T$ is the temperature. Below we
will assume that the $I-V$ characteristic of the boundary between two
metals is similar to that of a diode.  The high power (or strongly
nonlinear) regime corresponds to the condition $eV \gg kT$. 

\vbox{
\vspace{-0.0 in}
\hbox{
\hspace{-0.3in} 
\epsfxsize 4.0 in \epsfbox{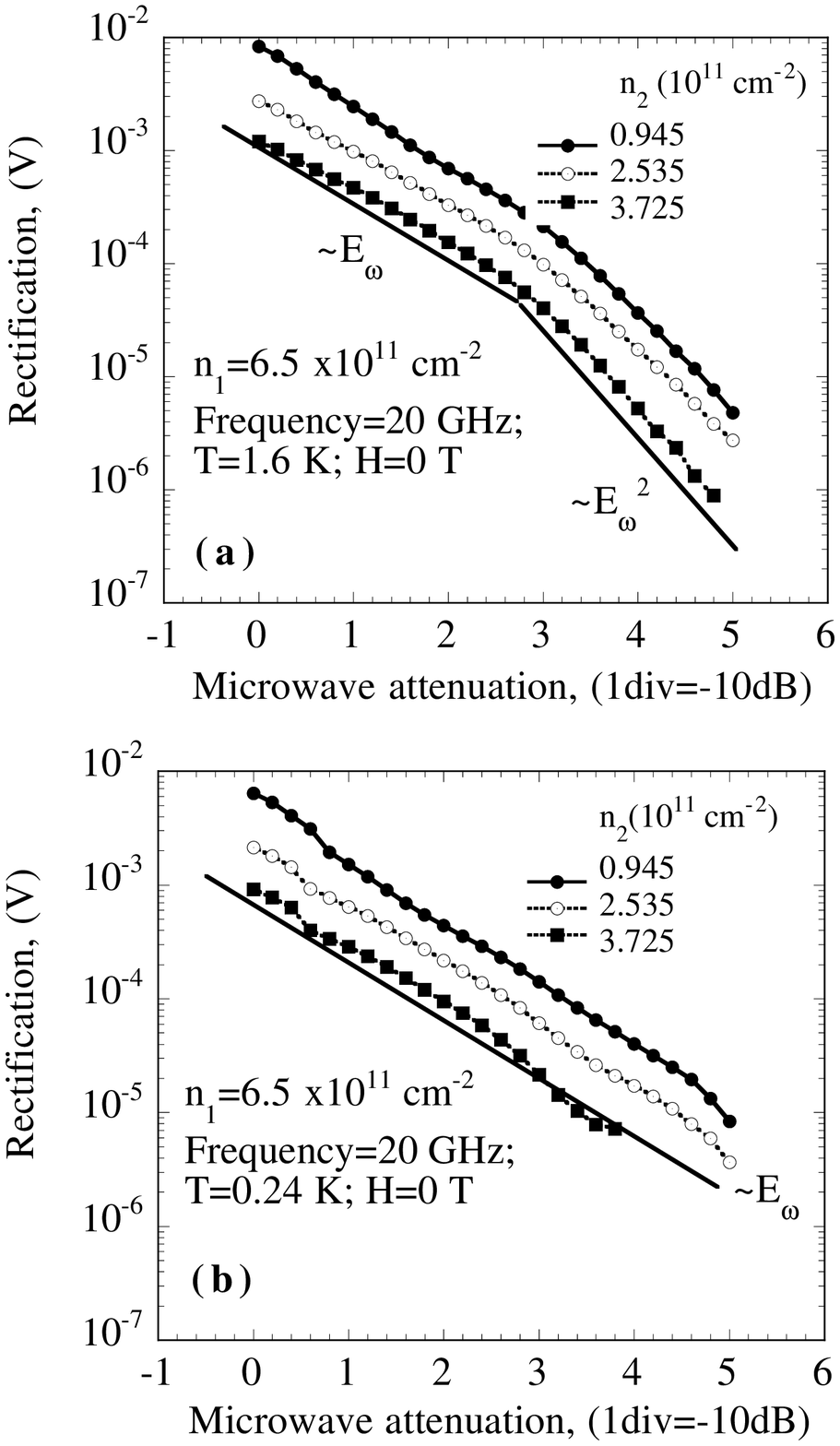} 
}
}
\vskip 0cm
\refstepcounter{figure}
\parbox[b]{3.3in}{\baselineskip=12pt FIG.~\thefigure.
a) Rectified signal versus microwave attenuation on a logarithmic scale.
One division of the horizontal axis represents a $-10$ dB change of the
microwave intensity.  The curves are taken for three different values  of
electron density of 2D-metal \#2  and fixed electron density of
2D-metal \#1, as shown.  The two solid curves represent linear
dependence and quadratic dependence of the rectification on the amplitude
of the  microwave field $E_\omega$.  (a) At T=1.6K the 
dependence of the rectified signal on microwave amplitude changes from
quadratic (perturbative regime) to linear (strong nonlinear regime). b)
At T=0.24K only linear dependence on the microwave amplitude is
observed (strongly nonlinear regime).
\vspace{0.10in}
}
\label{4} 
 In the
presence of the microwave field, the voltage drop across  the boundary is
$V=V_{\omega}+V_{dc}$, where $V_{\omega}=E_\omega d$ ($d$
is the
thickness of the boundary) is the voltage due to the microwave radiation
and $V_{dc}$ is the measured dc rectified voltage.  In order to yield zero
dc current, the dc voltage $V_{dc}$ must be comparable with $V_{\omega}$
at $eV \gg kT$.  This follows directly from an
estimate of the average (dc) current $<I(V)>=0$ in the strongly nonlinear
regime, ($eV \gg kT$), using the diode $I-V$ curve (see Eq. (1)).  This
conclusion should be valid qualitatively for any other strongly nonlinear
$I-V$ curves.  Thus, the magnitude of the microwave voltage drop across
the boundary between two metals should be of the order of the
rectified signal in the high power regime: $V_\omega
\approx V_{dc}$. 

According to Eq. (1) the crossover to the perturbative regime ($V_{dc}
\sim E_{\omega}^2$) should appear at $eV_\omega \approx kT$, which gives
$eV_{dc} \approx kT $ at the crossover.  The rectified voltage  $V_{dc}
\approx 10^{-4}V$ at the crossover (see Fig. 4 (a)). Therefore the
crossover temperature is $T_{cross}\approx eV_{dc}/k=1.2$ K, which
is of the same order as the temperature of the experiment - $T=1.6$ K (see
Fig. 4 (a)).

\subsection{Temperature dependence of rectification}
            
As shown in Fig. 4, the rectified signal $V_{dc}$ is a weak function of
temperature in the strongly nonlinear regime.  In the perturbative
nonlinear regime ($V_{dc} \sim E_{\omega}^2$) the microwave rectification
increases substantially with decreasing temperature $T$.  Fig. 5 shows
the temperature dependence of the rectified signal at different electron
densities, as labeled. 
\vbox{
\vspace{0 in}
\hbox{
\hspace{ -0.30 in} 
\epsfxsize 3.8 in \epsfbox{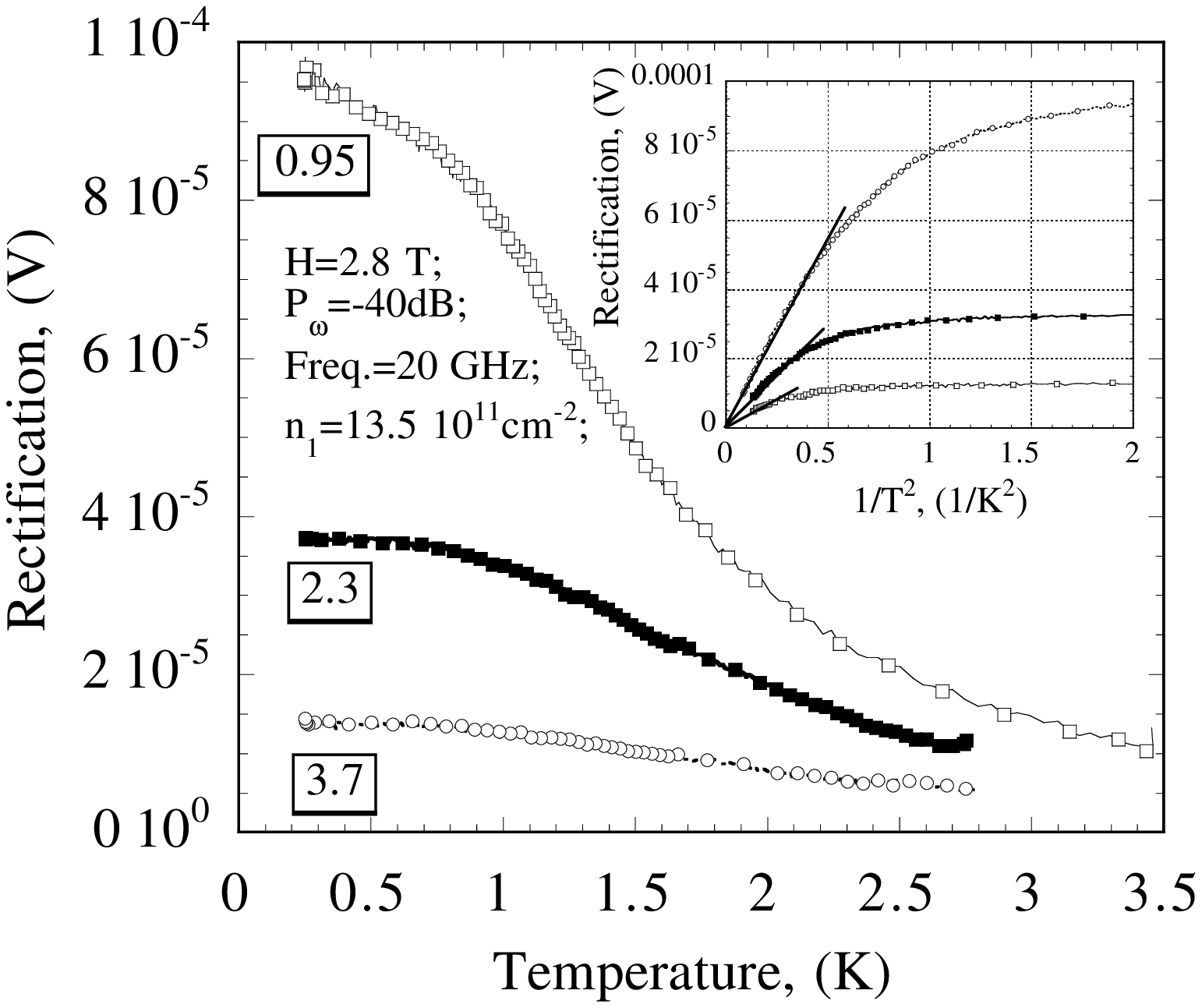} 
}
}
\vskip -3cm
\refstepcounter{figure}
\parbox[b]{3.3in}{\baselineskip=12pt FIG.~\thefigure.
Rectification versus temperature.  The electron density of 2D-metal 
\#1 is kept constant at $n_1=13.5x10^{11}$ cm$^{-2}$.  The electron
density of 2D-metal \#2 is different for different curves as labeled in
units of $10^{11}$ cm$^{-2}$. The magnetic field is $2.8$ Tesla.  The
insert shows the same data plotted versus $1/T^2$.
\vspace{0.10in}
}
\label{5} 

\vbox{
\vspace{0 in}
\hbox{
\hspace{0 in} 
\epsfxsize 3.6 in \epsfbox{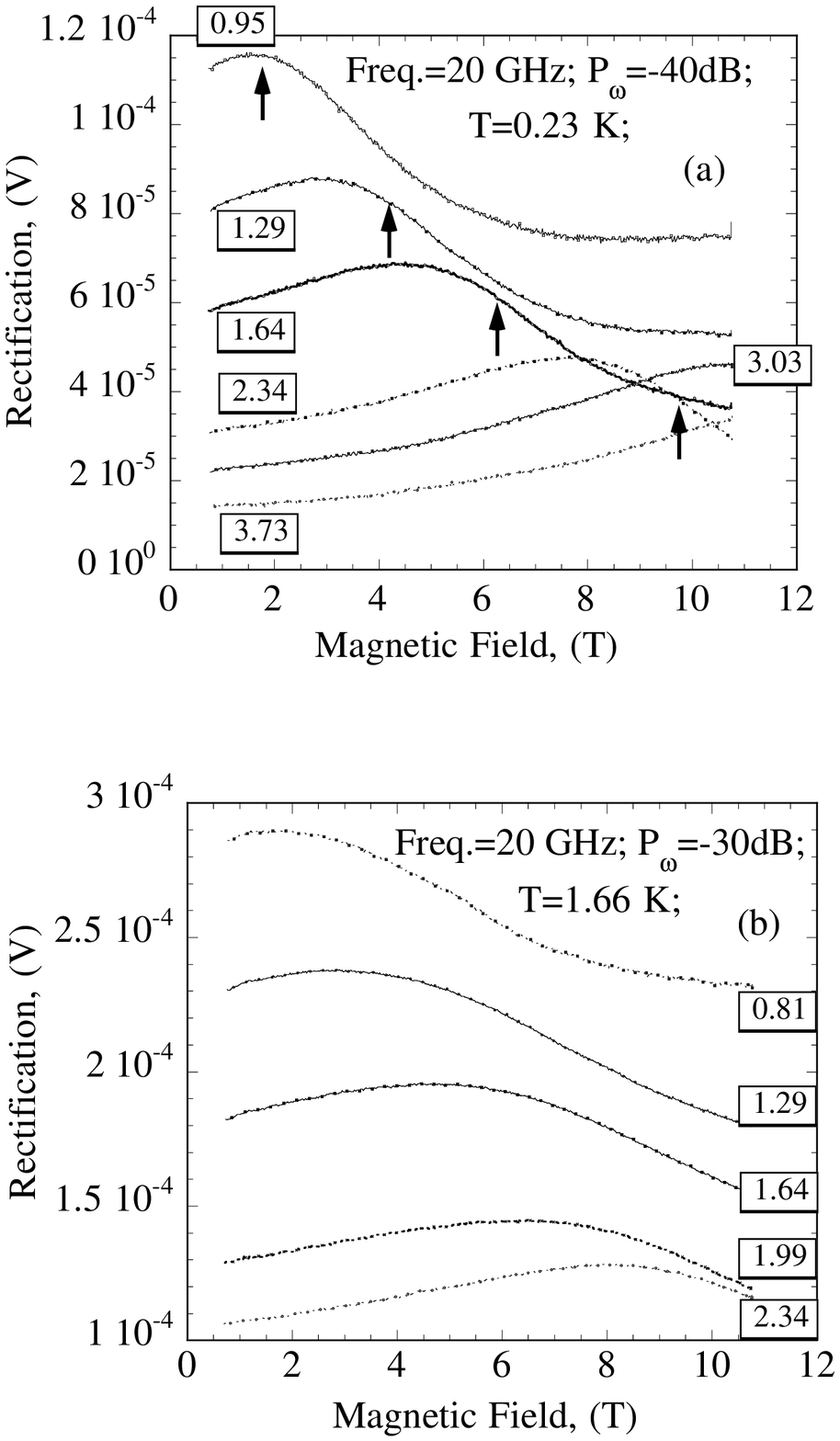} 
}
}
\vskip 0.5cm
\refstepcounter{figure}
\parbox[b]{3.3in}{\baselineskip=12pt FIG.~\thefigure.
(a) Rectification versus in-plane magnetic field at $T=0.23$ K. 
Curves are taken for different fixed electron densities in the main area.
Values of electron density in units of 10$^{11}$ cm$^{-2}$ are labeled.
The arrows  indicate the magnetic  fields
corresponding to complete spin  polarization of the electrons under the main
gate\cite{vitkalov_ferro}.  (b) Rectification versus
in-plane magnetic field at $T=1.66$ K.  The direction of the magnetic field is
parallel to the boundary between the two metals.
 
\vspace{0.10in}
}
\label{6}    
The data were obtained at finite magnetic field to avoid a change
(about 20\%- 30\%) of the microwave field $E_\omega$ near the sample due
to the superconducting transition of the Al gate at $T_c=1.4$ K. The
increase of the nonlinearity with decreasing electron temperature
indicates that the conductivity itself is not the relevant parameter
responsible for the rectification.  The electrical charge accumulation
near the boundary due to the microwave field, one of the possible
mechanisms responsible for the rectification, should be less important at
lower temperature due to the higher conductivity of the dilute 2D system
at low $T$.  

The saturation of the temperature dependence at low temperature may be
related partially to overheating of the electrons by the microwave
radiation as well as to the transition from the regime where $V_{dc} \sim
E_\omega ^2$ to the regime where $V_{dc} \sim E_\omega $ (see
Fig. 4).  The insert to Fig. 5 demonstrates the high temperature behavior
of the signal. At high temperature the rectification is proportional to
$1/T^2$ at different electron densities. The $T^{-2}$ dependence of the 
rectification is different from the $1/T$ dependence expected for a
regular diode.  Using the parameter $eV/kT=(V_{dc}+V_\omega)/kT$ as a
small perturbation in Eq. (1), the rectified signal was
found to be: $V_{dc} \sim V_\omega(eV_\omega/kT)$.

\subsection{In-plane magnetic field dependence of rectification}   
    
The dependence of the rectified signal on in-plane magnetic field is
shown in Fig. 6 for different electron densities at two different
temperatures.  The contact gate area was kept at high electron density
$n_{1}=13.5 \times 10^{11}$ cm$^{-2}$. Different curves correspond to
different electron density under the main gate, as labeled.  The direction of
the magnetic field is parallel to the boundary between the two metals.
The rectification is a nonmonotonic function of in-plane magnetic field. 
The arrows shown in Fig. 6 indicate the magnetic  fields corresponding to
complete spin polarization of the electrons under the main
gate\cite{vitkalov_ferro}.  The magnetic field of about
$12$ T was not sufficient to fully polarize the high density of electrons
under the contact gate \cite{vitkalovangular}.

\section{Discussion}
The mechanism that gives rise to the observed rectification is unknown and
is the subject of current investigation.  Although they are expected to be
important, it is currently unclear how to include the effect of
electron-electron interactions.  In this section we first consider a
simplified model of non-interacting electrons and show that these simple
considerations do not account for our observations.  We then speculate
about other possible explanations.
 
The observed rectified signal depends on the difference between electron
densities in the two 2D layers (see Fig.2).  Almost perfect symmetry of
the effect with respect to exchange of the gate voltages (see Fig.3)
indicates that the rectification is generated near the boundary
separating the two 2D electron systems.  It is well known
\cite{Landau}, that there is an internal electric field near the boundary
separating two metals.  This electric field creates a contact difference
of potentials $\Delta \phi$. The difference in contact potentials is
equal to the difference of the work functions or chemical potentials
$\mu_{1,2}$ of the metals considered independently (in other words, when
the metals are not connected to each other).  At thermodynamic
equilibrium the chemical potential has to be the same when the system is
connected. Thus, the potential difference $\Delta
\phi=\mu_2 - \mu_1$ should be generated near the boundary.   

Let us assume that the contact difference of potentials $\Delta
\phi=\mu_2 - \mu_1$ is the main reason for the rectification near the
boundary between the two metals.  This suggestion is completely consistent
with the symmetry of the effect with respect to exchange of the
gate voltages (see Fig.3).  Below we estimate the rectification of the
microwave radiation near the conducting boundary between two normal 2D
metals using the Boltzmann equation.

To simplify the following calculations we introduce a model where a pair
of 2D metals contiguous through the conducting boundary is replaced by a
single metal of strongly inhomogeneous electron density (see Fig. 1 (c)). 
The inhomogeneity along the direction perpendicular to the boundary
between the original 2D metals is created by the voltages applied to
the gates.  We choose a coordinate system where the boundary
lies along ``y'' axis.  Then both the electron density and the chemical
potential depend on ``x''
$ (\mu (x) = \frac{n(x)}{D_0},$ where $ D_0 $ is the electron density  
of states at the Fermi surface).
                              
The Boltzman transport equation for the electron distribution function
$f(x, {\bf p},t) $ is taken within the relaxation time approximation:
    $$
\frac{\partial f}{\partial t} + v_x \frac{\partial f}{\partial x} +
\frac{\partial \bf p}{\partial t} \frac{\partial f}{\partial \bf p} =   
- \frac{f - f_{eq} }{\tau}
                          \eqno(2)$$
where the function $ f_{eq}$ describes the  equilibrium state of the
electron system in the absence of the microwave radiation $ E(t)$ and
equals the Fermi distribution function:
    $$
f_{eq} = f_0 (\varepsilon, \mu (x), T).
                              \eqno(3)$$
Following [9] we expand $ f(x, {\bf p}, t)$ in harmonic polynomials,
keeping the first three terms of the expansion. Such an approximation
is justified for a weak microwave field because the next terms of the
expansion give smaller corrections in terms of the microwave field
magnitude.  Substituting the expansion into the Boltzman equation we get a
set of equations:
  $$
\frac{\partial f_1^\alpha}{\partial t} +
\frac{e}{m}E_\alpha(t) \frac{\partial f_{eq}}{\partial \varepsilon} +
2e \bigg (E_\beta (t) + \frac{1}{e} \frac{\partial \mu}{\partial x}
\delta_{\beta x} \bigg ) f_2^{\beta \alpha}
                                            $$
  $$
- e \frac{p^2}{2m} E_\alpha (t)
\frac{\partial f_2^{\alpha \beta}}{\partial \varepsilon}
\delta_{\alpha \beta} = - \frac{f_1^\alpha}{\tau};
                                     \eqno(4)$$

  $$
\frac{\partial f_2^{\alpha\beta}}{\partial t} +
\frac{e}{m}E_\alpha(t) \frac{\partial f_{1}^\beta}{\partial t} =
- \frac{f_2^{\alpha \beta}}{\tau}.
                                  \eqno(5)$$
 Here $f_1^\alpha, f_2^{\alpha\beta}$ are  the coefficients of the
expansion of the electron distribution function in harmonic polynomials;
$ m $ is the electron effective mass, $ \alpha ,\beta = x,y.$

The term $\frac{1}{e} \frac{\partial \mu}{\partial x} \delta_{\beta x}$ in
(4) describes the internal ``electrochemical'' field arising from the
inhomogeneity of the electron density. In the absence of the time
dependent field $ E (t) $, this term balances the external field produced
by the gate voltages and provides equilibrium of the electron system.

Solving Eqs. (4), (5), we can calculate the dc contribution to the
electron current density.  This rectified current is generated
perpendicularly to the boundary (along the ``x'' direction).  In
the present experiments, where the microwave frequency $\omega$ is
smaller than the effective collision frequency $(\omega \tau < 1) $ we
arrive at the simple expression for the rectified current:
  $$
j_{dc} = - \frac{4 e^3 \tau^3}{\pi m^2}
\frac{\partial n}{\partial x} E_\omega^2,
                                     \eqno(6)$$
 which enables us to estimate $ V_{dc}$ as follows:
  $$
 |V_{dc}| = \left | \int_{- \infty}^\infty \frac{j_{dc}(x)}{\sigma_0(x)}dx
\right |.
                                     \eqno(7)$$
 Here $\sigma_0 (x)$ is the Drude conductivity  corresponding to electron
density $ n(x) $ and relaxation time $ \tau (x).$

It was observed in the present experiments that when the minimum $ n(x)$
is larger than a certain value $ n_0 \; (n_0\approx 2.5 \times 10^{11}$
cm$^{-2}$) the relaxation time $\tau $ is nearly independent on the
electron concentration and can be treated as a constant in carrying out
the integration over ``x''in (7).  Using this approximation we get:
  $$
|V_{dc}| = \frac{4}{\pi} \frac{e}{m} \tau^2 \ln \frac{n_1}{n_2}E_\omega^2.
                                     \eqno(8)$$

These results based on the Boltzman transport equation are in agreement
with the experiments in several important respects.  First, it follows
from (6)--(8) that rectification can take place only when there exists a
difference in the electron densities of the 2D metallic regions
separated by a conducting boundary.  Secondly, the above considerations
give a correct relation between $ V_{dc}$ and $E_{\omega}$ for small
power levels of the microwave field $ (V_{dc} \sim E_\omega^2)$.

There are, however some discrepancies between the theoretical model
and the experiments. Using the thickness of the boundary $d \approx 0.1 \mu$, the electric field near the boundary is estimated to be $ E_\omega \approx V_{dc}/d= 10^3 \frac{V}{m}$ at microwave attenuation -30dB and the temperature $T=1.6K$ (see fig. 4(a)). Using the estimates $ m \approx 0.2 m_0 \, (m_0 $ is the mass of a
free electron), $ \tau \approx 3 \times 10^{-12} s^{-1}$ and eq.(8), we obtain 
$ V_{dc}^{est} \sim 10 \mu V,$ which is less than the rectification 
$ V_{dc} \approx 100 \mu$ observed at -30dB at $T=1.6KT$ experimentally (see fig.4(a)).
Moreover the perturbative parameter used in the Boltzman
approach is proportional to the ratio of microwave electric potential
$V_\omega \approx V_{dc} \approx 10^{-5}$ V at $-40$ dB (see Fig.4b) to
the contact difference of potentials  $\Delta \phi=\mu_2 - \mu_1 \approx
4$ mV formed at the boundary between two metals.  This ratio is
about $10^{-2}$ in our experiment at $P_\omega$=-40 dB, which 
according to the above approach indicates we're in the perturbative regime
of rectification: $V_{dc} \sim E_\omega ^2 $ (see  eq.(8)). However this
contradicts our experimental observations at these microwave power
levels (see Fig.4b): 
$V_{dc} \sim E_\omega$ at $n \approx 1 \times 10^{11}$ $cm^{-2}$.  

The magnetic field dependence of the rectification exhibits
nonmonotonic behavior.  This behavior can be qualitatively understood in
terms of the competition between two effects.  The dependence of the
chemical potential of the electrons with magnetic field is given by
$\Delta \mu=-1/2(d\chi/dn) \times H^2$, where we have assumed that
$(d\chi/dn)$ is independent of magnetic field. The derivative of
the magnetic susceptibility with respect to electron density,
$(d\chi/dn)$, is negative due to the decrease of the magnetic
susceptibility $\chi(n)$ with increasing electron density
\cite{Shashkin,pudalov_SdH}.  The increase of the chemical potential
appears to be stronger at low electron density in Si-MOSFET's.  This
implies that the difference of the chemical potentials $\mu_2-\mu_1$ is
reduced in a magnetic field, decreasing the rectification.  On the other
hand, the conductivity $\sigma$ is also reduced, with a consequent
increase in the rectification (see eq.(7)).  Careful analysis of these
competing effects is needed in order to understand the dependence of the
rectification on magnetic field. 

The strong nonlinearity observed in these experiments appears to be 
similar to the response of an ordinary diode.  It is well known that the
diode rectification is proportional to the amplitude of the electric
field in the high power regime.  In the case of an ordinary diode, the
rectification is due to the potential barrier formed between the $p$ and
$n$ regions, which induces an exponential dependence of the current on
applied voltage \cite{Levine}.  Another important property of the diode 
is the presence of two kinds of carriers: electron and holes. 
Below we will consider these possibilities.  

One of the reasons for the potential barrier is simple electrostatics.
The barrier could be the result of the finite width $d$ of
the split  between two gates.  When the voltages of the gates are equal,
then a drop in density will occur under the split.  However the width of
the split, $d=50-70$ nm, is smaller than the distance between the 2D
electrons and the gates $d_{ox}=150$ nm.  Therefore, the decrease in
density is expected to be quite small. Moreover at large differences
between the gate voltages (an order of magnitude in our
experiment) there is a very sharp drop in the electron density.  In this
case one expects a monotonic change in electron density between the two
metals \cite{heemskerk}.  We must also note that even in the
presence of such an electrostatic potential barrier the rectification
should be absent at low microwave voltages \cite{simmons}.      
 
An additional potential barrier between two dilute metals could also
arise due to the  presence of different charged excitations in a 2D dilute
system on a Si surface, as proposed in several papers
\cite{Kirkpatrick,Phillips,Chakravarty,Chamon}.  For example,  if the
majority carriers in one metal are Fermi particles while the other 2D
metal contains mostly Bose particles (for example paired electrons) with
an energy of dissociation $\Delta$, then an additional potential barrier
can, in principle, be formed between the two metals.  The origin of such
an electrochemical barrier is the reaction of formation of the Bose
particles $e+e=2e$ (or other correlated structure).   A similar
electrochemical barrier will be formed also between two metals  with
different strength of the electron-electron interaction.  The $1/T^2$
temperature tail of the rectification (see fig. 5) indicates that the
energy scale responsible for the rectification effect is proportional to
the $T^2$. This energy scale has the same temperature dependence as the
rate of electron-electron scattering in a metal. 
 
Finally, the observed rectification could be the result of variation of
the density of electron states with energy.  However, the very strong
nonlinear response observed in our experiments would require that the
density of electron states be a very strong function of energy.

\section{Conclusion}
In summary, we report the observation of the rectification of microwave
radiation by the boundary between two two-dimensional metals on the
surface of Si.  The effect was investigated as a function of electron
density, microwave power, temperature, and in-plane magnetic field.  The
rectified voltage depends on the densities of the two metals, and goes to
zero when the electron densities are equal.  At the lowest measured
temperature $T=0.23$ K the rectified signal is directly proportional to
the amplitude of the microwave field if one of the metals is close 
to the quantum phase transition \cite{vitkalov_ferro}, while the other is
kept at high electron density.  This signals a strongly nonlinear regime
even when the contact potential difference between the metals is much
larger than the potential associated with the microwave field.

S. A. V. thanks M. Reznikov and B. Altshuler for usefull comments. 
This work was supported by DOE grant No.DOE-FG02-84-ER45153.   
Partial support was also provided by NSF grant
DMR-9803440.

\pagebreak

\end{multicols}


\begin{references}
\bibitem{high_dens} The small ($<10 \mu V$ rectification signal (not shown in the fig.4a) at high density is comparable with an offset voltage at $T=0.24K$. 
It needs an additional study.
\end{references}
\end{document}